\documentclass[preprint,prd,showpacs]{revtex4}
\usepackage{graphicx}
\usepackage{epsfig}
\usepackage{subfigure}
\usepackage{bm}
\usepackage{multirow} 
\def\d{{\rm d}}

\begin{document}

\title{Tau contamination in the platinum channel at neutrino factories}

\author{Rupak Dutta}
\email{rupak@imsc.res.in}      
\author{D.~Indumathi}
\email{indu@imsc.res.in}
\author{Nita Sinha}   
\email{nita@imsc.res.in}
\affiliation{%
The Institute of Mathematical Sciences, 
Chennai 600113, India
}
\preprint{IMSc/2011/3/1}

\begin{abstract}
  The platinum channel ($\nu_e$ or $\overline{\nu}_e$ appearance) has
  been proposed at neutrino factories as an additional channel that
  could help in lifting degeneracies and improving sensitivities to
  neutrino oscillation parameters, viz., $\theta_{13}$, $\delta_{CP}$,
  mass hierarchy, deviation of $\theta_{23}$ from maximality
  and its octant. This channel corresponds to $\nu_\mu \to \nu_e$
  ($\overline{\nu}_\mu \to \overline{\nu}_e$) oscillations of the initial
  neutrino flux, with the subsequent detection of electrons (positrons)
  from charged current interactions of the $\nu_e$ ($\overline{\nu}_e$)
  in the detector. For small values of $\theta_{13}$, the dominant
  $\nu_\mu \to \nu_\tau$ ($\overline{\nu}_\mu \to \overline{\nu}_\tau)$
  oscillation results in this signal being swamped by electrons arising
  from the leptonic decay of taus produced in charge-current interactions
  of $\nu_\tau$ ($\overline{\nu}_\tau$) with the detector. We examine for
  the first time the role of this tau contamination to the electron events
  sample and find that it plays a significant role in the platinum channel
  compared to other channels, not only at high energy neutrino factories
  but surprisingly even at low energy neutrino factories. Even when the
  platinum channel is considered in combination with other channels such
  as the golden (muon appearance) or muon disappearance channel, the tau
  contamination results in a loss in precision of the measured parameters.


\end{abstract}


\maketitle
\section{Introduction}

In the last decade, evidence for neutrino oscillations has been obtained
from experiments with various neutrino sources. The parameters that
characterize neutrino oscillations within a three-flavor neutrino mixing
framework are the two mass squared differences $\Delta m_{21}^2$ and
$\Delta m_{31}^2$ ($\Delta m_{ij}^2 = m_i^2 - m_j^2$), the three mixing
angles $\theta_{12}$, $\theta_{13}$, and $\theta_{23}$, and the Dirac
phase $\delta_{CP}$. While the parameters $\theta_{12}$, $\theta_{23}$,
$\Delta m_{21}^2$ and the magnitude of $\Delta m_{31}^2$ are relatively
well known, the sign of $\Delta m_{31}^2$ and hence the neutrino mass
hierarchy, as well as the across-generation mixing angle $\theta_{13}$
are unknown, the latter having just an upper bound \cite{chooz}. The
hardest to measure will be the CP phase.  Many experiments are set
to measure one or more of these parameters, with new upcoming super-beam
experiments proposing precision measurements of various oscillation
parameters although these experiments may not be sufficient to determine
all the sub-leading parameters accurately, particularly for all fractions
of the CP violating phase $\delta_{CP}$.

Neutrino factories have been proposed to provide neutrino beams from
future muon storage rings where the muons decay in long
straight sections, producing both muon- and electron-type neutrinos (and
anti-neutrinos). Neutrino factories have been mooted as an excellent
set up for precision measurement of neutrino oscillation parameters
when the across-generation mixing angle $\theta_{13}$ is small,
$\sin^22\theta_{13} \lesssim 10^{-2}$ \cite{ids-nf}. These future
facilities have the advantage of high neutrino fluxes with suppressed
beam backgrounds and can provide baselines from about 1000 km for
low energy neutrino factories (LENF) to baselines as long as the
magic baseline $\sim 7500$ km for the high energy neutrino factories
(HENF). Precision measurements of $\theta_{13}$, leptonic CP violation,
the type of neutrino mass hierarchy, i.e, the sign of $\Delta m_{31}^2$,
deviation of $\theta_{23}$ from $\pi/4$ (maximal) and if found, its
octant, may be feasible only at neutrino factories.

The key to this high sensitivity is the clean separation of {\em
wrong sign} (WS) events from the right sign (RS) ones by lepton charge
identification. For instance, if the source of the neutrino beam in a
muon storage ring is $\mu^-$, then the beam is a (precisely known)
mixture of $\nu_\mu$ and $\overline{\nu}_e$.  In the absence of
oscillations, we would expect to see muons (from charged current (CC)
interactions of $\nu_\mu$) at a far detector; detection of anti-muons, or
WS leptons, would then be an unambiguous signal of neutrino
oscillations (via $\overline{\nu}_e \to \overline{\nu}_\mu \to \mu^+$,
the last step occurring during interactions with the detector). Such a
WS muon (or appearance) signal from $\nu_e \to \nu_\mu$ oscillations,
called the {\em `golden channel'}, has been studied extensively as it is
sensitive to all the above mentioned oscillation parameters.

There are, however, correlations and degeneracies which in turn
deteriorate the achievable precision. Even with neutrino and anti-neutrino
running, there is an eightfold ambiguity~\cite{deg} due to the intrinsic
$(\delta_{CP}, \theta_{13})$ degeneracy, the unknown sign of $\Delta
m^2_{31}$ and the unknown octant of $\theta_{23}$.

These correlations and degeneracies can be reduced \cite{whichlenf,HA}
by improving the statistics or including two other appearance channels
\cite{huber}: {\em `silver'} ($\nu_e \to \nu_{\tau}$) and {\em `platinum'}
($\nu_{\mu} \to \nu_e$). To date, the design of a detector capable
of a large sample of CC $\nu_\tau$ or $\overline{\nu}_\tau$ has been
a challenge. Hence in this study we assume that the neutrino factory
set up has detectors with capability of muon as well as electron charge
identification; however no tau detection capability is present. Detection
of electrons (positrons) at a far detector for a $\mu^-$ ($\mu^+$) beam,
called appearance or wrong sign electrons, would then be an unambiguous
signal of the platinum channel.

In the absence of a detector capable of identifying $\nu_\tau$ or
$\overline{\nu}_\tau$, the $\nu_\tau$'s from
$\nu_e\rightarrow\nu_\tau$ and $\nu_\mu\rightarrow\nu_\tau$
oscillations will produce taus from CC interactions in the
detector, which can subsequently decay to muons. These muon events
will not be distinguishable from the `direct' muons (i.e., those
produced from $\nu_\mu$ CC interactions) and hence will add to the
golden channel and muon disappearance events, respectively. This tau
contamination in the disappearance channel has been discussed in
detail in Ref.~\cite{IN}. The issue of the tau contribution to the
golden channel muon sample was also briefly mentioned in
Ref.~\cite{IN}, but the detailed analysis and quantitative results
were presented in Ref.~\cite{Donini}.

In this note, we highlight the hitherto neglected electron events
from $\nu_\mu\to \nu_\tau$ oscillations with leptonic decay of taus
produced in CC interactions in the detector to electrons; these add to
the usual or direct `platinum' channel, i.e., $\nu_{\mu} \to \nu_{e}$
oscillations with electrons produced in CC interactions of $\nu_e$'s with
the detector. Since $P_{\mu\tau}\gg P_{\mu e}$, this tau contribution
to WS electron (appearance) events will in fact be larger than
the direct electron appearance events. This is in great contrast to the
case of the muons where the tau contributes dominantly to the {\em right
sign} muon signal. It is expected therefore that this tau contribution
has a stronger impact on the electron appearance channel and hence
the sensitivity of this channel to the various neutrino oscillation
parameters; this additional contribution to the WS electron events
in neutrino factories has not been discussed so far. In particular,
we focus on how this large tau contamination to the electron events
can alter the sensitivity to the mixing angle $\theta_{13}$ and the CP
violating phase $\delta_{CP}$.

The paper is organized as follows. In Section~\ref{NM}, we briefly
describe details of the neutrino factory set up and show typical rates at
typical detectors of choice. In Section~\ref{RD}, we report our results for
various processes and their sensitivity to the yet unknown mixing angle
$\theta_{13}$ and $\delta_{CP}$. We conclude in Section~\ref{CON}.

 
\section{Detector and beam set up}
\label{NM}

The reach of various neutrino factory set ups with respect to various
neutrino oscillation parameters has been discussed in many papers in great
detail \cite{nfgeneric}. Here we examine two cases, one a high-energy
neutrino factory (HENF) from muons at 25 GeV as in Ref.~\cite{huber},
and a low energy neutrino factory (LENF) set up with a muon source at
4.5 GeV with a totally active scintillator detector (TASD) or a liquid
argon detector (LAr) as in Ref.~\cite{Li}.

The event rates for production of leptons of flavor $j$ in the detector
from `direct' and `tau-induced' channels are defined as
\begin{eqnarray} \nonumber
{\cal {N}}^{direct}_{ij} & = & \kappa  \int \Phi_i P_{ij} \sigma_j (\nu_j
\to j) \epsilon_j~,  \\
{\cal {N}}^{\tau}_{ij} & = & \kappa  \int \Phi_i P_{i\tau} \sigma_\tau
(\nu_\tau \to \tau) \Gamma (\tau \to j) \epsilon_j ~, 
\end{eqnarray}
where $\kappa$ accounts for the exposure (size of detector and years of
running), the initial flux ($\Phi \equiv \d^2\Phi/\d E\d\cos\theta$)
corresponds to $i = e, \mu$ type neutrinos, the differential cross-section
($\sigma \equiv \d^2\sigma/\d E \d\cos\theta$) is for CC interactions
(quasi-elastic, resonance and deep inelastic processes) producing
lepton $j$ or $\tau$ in the detector and $\epsilon_j$ are the detection
efficiencies. Note that the oscillation probability $P_{ij}$ is a function of
both $(E,\cos\theta)$ of the neutrino. In addition, for the $\tau$-induced
channel, we include the decay rate, $\Gamma \equiv \d^2\Gamma/\d E\d
\cos\theta$ for $\tau \to j$ where $j = e$ is the case of interest
here. The integration is over all the relevant variables, including
resolution functions, corresponding to bins in the observed lepton energy
and direction ($E_j^{\rm obs}, \cos\theta_j^{\rm obs})$. For the
analysis, we have added the events from both $\mu^-$ and $\mu^+$
beams. For details of computation of the kinematics, cross section,
and flux, we refer to Ref.~\cite{IN}. Here we merely highlight four
different sets of events: $(ij)=(\mu e), (e\mu)$ corresponding to WS
electron and muon (appearance) channels respectively, and $(ij) = (ee),
(\mu\mu)$ corresponding to RS electron and muon (disappearance) channels
respectively. When the process occurs via $\tau$ production and decay,
it simply adds to one of these channels and is labelled as a
`tau-induced' appearance or disappearance event.

While the $\Phi_e$ flux gives rise to RS electron events in the detector,
the $\Phi_\mu$ flux gives rise to WS events. Hence, while $\tau$-induced
events contribute to both the RS and WS events, they contribute dominantly
to the WS contribution since the $\nu_{\mu} \to\nu_{\tau}$ oscillation
probability is driven by a nearly maximal mixing angle $\theta_{23}$ so
that $P_{\mu\tau} \gg P_{e\tau}$.

Despite the kinematic suppression of the CC cross section for tau
production due to the large tau mass, there is a sizeable amount of tau
production rate above threshold ($E_\nu^{thr}\sim 3.4$ GeV). Even with
the total tau decay rate into electrons of about $17\%$, these tau induced
events are substantially larger than the direct events, particularly in
the low electron energy bins. This turns out to be true not only for a
HENF of 25 GeV, but surprisingly even for an LENF of 4.5 GeV. Hence, tau
contamination must be taken into account while analyzing the platinum
channel at any neutrino factory.

While analyzing the effects of this tau contamination we use typical
oscillation parameters, $\Delta m^2=2.4\times 10^{-3}$ eV$^2$,
$\theta_{23}=45^\circ$, $\sin^2 \theta_{12}= 0.304$ and $\Delta
m_{21}^2=7.65\times 10^{-5}$ eV$^2$ (we use the symmetric notation:
$\Delta m^2\equiv m_3^2-(m_1^2+m_2^2)/2$). We analyze the events for
sensitivity to $\theta_{13}$ and $\delta_{CP}$ for a small input value
of $\theta_{13}$, $\theta_{13}=1^\circ$. A threshold of 0.5 GeV for
electron detection is used. We use an overall normalization error of 0.1\%
for direct electron events. However, for the total (direct+tau) events,
due to the larger uncertainties in the tau production cross-section, a
larger 2\% error is used. (Note that the presence of a near detector will
not help reduce the uncertainties in the CC tau production cross-section
and this forms an important factor in limiting the precision measurement
of neutrino oscillation parameters). We deal with a HENF and LENF in turn.

\section{Sensitivity to neutrino oscillation parameters $\theta_{13}$
and $\delta_{CP}$}
\label{RD}

\subsection{Results for 25 GeV HENF}
\label{henf}

\begin{figure}[htb]
\begin{center}
\includegraphics[width=10cm]{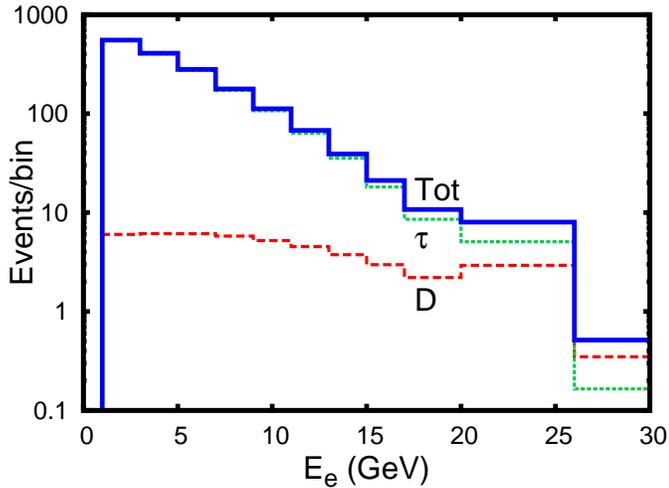}
\end{center}
\caption{Electron event rates as a function of observed electron energy
at a HENF with $25\,\mbox{GeV}$ $\mu^\pm$ beams (both polarities) over a
5 year exposure with a 15 kton detector located at a baseline of
$L=4000$~km.}
\label{fig:henfevents}
\end{figure}

For our analysis we assume detector characteristics as specified
in Ref.~\cite{huber}. The neutrino beam interacts with a 15 kton
detector~\cite{huber} located at a distance $L=4000$ km from the
source. At this baseline there is sensitivity to $\delta_{CP}$ as well
as $\theta_{13}$. We assume $5.0\times 10^{20}$ useful muons per year,
per polarity, and a running time of 5 years. The energy resolution
of electrons is taken to be 15\% and efficiency is 20\%. It can be
seen from Fig.~\ref{fig:henfevents} that the tau contribution to the
electron events (marked $\tau$ in the figure) is substantial and can in
principle alter the sensitivity to the oscillation parameters. Neglect
of the tau contribution will lead to measuring incorrect values of these
parameters. The magnitude of the direct (tau-induced) events is driven by
the small $\theta_{13}$ (large $\theta_{23}$); the direct events (marked D
in the figure) will be more appreciable for larger $\theta_{13}$. However,
the shape difference between the two contributions will remain: note
that the shape of the direct events is almost flat at low energies,
while the number of events quickly falls with increasing energy for the
tau contribution) resulting in rather different sensitivity of the total
events (labelled ``Tot") to the parameters.

\begin{figure}[htp]
\begin{center}
\includegraphics[width=0.45\textwidth]{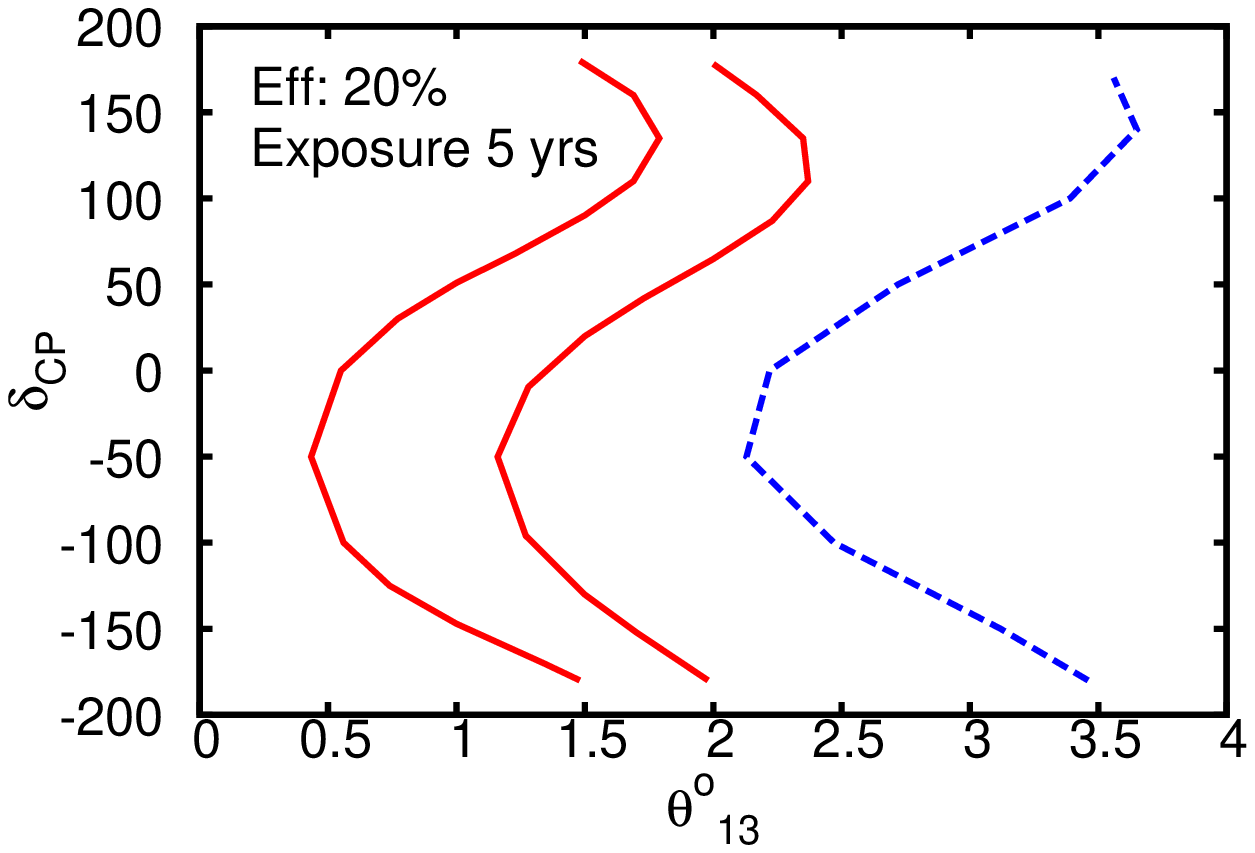}
\includegraphics[width=0.45\textwidth]{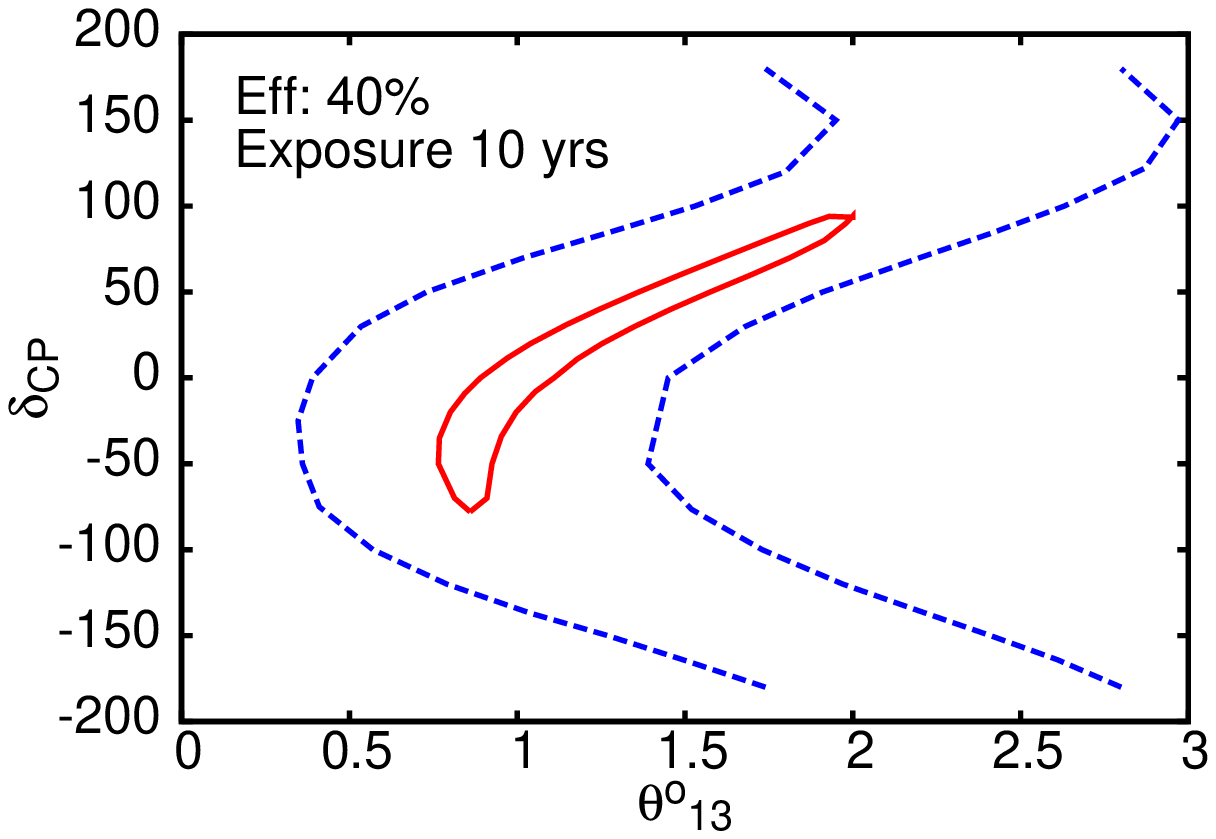}
\caption{(L) 99\% CL contours in $\theta_{13}$--$\delta_{CP}$ space at
a HENF with $25\,\mbox{GeV}$ $\mu^\pm$ beams (both polarities) over a
5 year exposure with a 15 kton detector located at a baseline of $L =
4000$ km. The region within the leftmost contours is the allowed space
when tau-induced events are neglected; including tau-induced events
dramatically worsens the sensitivity to these parameters, giving only
an upper bound on $\theta_{13}$ (rightmost curve). (R) Doubling the
efficiency and exposure as well as increasing the detector mass to 50
kton improves the sensitivity to $\theta_{13}$ but the sensitivities are
always worse than in the case when the tau contribution is neglected.}
\label{fig:henfsens}
\end{center}
\end{figure}

We generate allowed regions in $\theta_{13}$--$\delta_{CP}$ parameter
space, for input values of $(\theta_{13}, \delta_{CP}) = (1^\circ,
0^\circ)$, keeping the other parameters fixed at their present best-fit
values~\cite{bestfit} with the normal hierarchy for the 2--3 sector.
The 99\% CL contours obtained by minimizing the chi-squared with a pull
corresponding to the normalization uncertainties as specified in
Section~\ref{NM} are shown in Fig.~\ref{fig:henfsens}.

The region within the two leftmost contours is the parameter space allowed
as a result of neglecting tau contributions both in the ``data'' and
in the theoretical fits. When these are taken into account correctly,
only an upper bound (rightmost curve) is obtained: hence this input
value of $\theta_{13}$ cannot be discriminated from zero. A substantial
increase in exposure and detector size and characteristics is required
in order to regain sensitivity to this value of $\theta_{13}$, as can
be seen from the contours in the right side of this figure. Even then,
sensitivity to $\delta_{CP}$ is entirely lost for this choice of input
parameters when the tau contribution is included; in addition, sensitivity
to $\theta_{13}$ is substantially reduced as well.

Note that it is hard to find ``reasonable" contours
to fit the ``true" data if tau-induced events are not included in the
theoretical fits at all, within the normalization uncertainty. No other
uncertainties or backgrounds have been taken into account here (and
elsewhere in the paper) since we primarily wish to emphasize the change in
sensitivity with the inclusion of the tau-induced events.

\subsection{Results for 4.5 GeV LENF}
\label{lenf}

We now examine whether there is any effect of the tau contamination
for a low energy neutrino factory. With the tau production threshold at
3.4 GeV, one would naively expect the tau contamination not to play
any role in a LENF set up~\cite{HS}. The details of the performance of
a low energy neutrino factory with muon energy 4.5 GeV and its
excellent sensitivity to oscillation parameters such as $\theta_{13}$
and $\delta_{CP}$ for $\sin^22\theta_{13} > 10^{-4}$, and to the
mass hierarchy for $\sin^22\theta_{13} > 10^{-3}$, when tau events
are neglected, are given in Ref.~\cite{Li}.  We repeat these
calculations, including the tau contribution, for a 4.5 GeV LENF with
$1.4\times 10^{21}$ useful muon decays per year, per polarity, and a
running time of 10 years, with a baseline $L = 1300$ km.

Here we consider two different detectors whose characteristics are given
in Ref.~\cite{Li}: a magnetized 20 kton totally active scintillator
detector (TASD), with electron detection efficiency 37\% (47\%)
below (above) 1 GeV, and 10\% energy resolution. We also consider a
future possible 100 kton magnetized liquid argon detector (LAr) with a
substantially higher (80\%) electron detection efficiency and a somewhat
worse energy resolution of 20\% (except for quasi-elastic events, where
it is 5\%). We also use a larger normalization error of (5\% for direct
and 5.5\% for the total events), due to the higher systematic error
expected for LAr detectors.

\begin{figure}[htp]
\begin{center}
\includegraphics[width=0.45\textwidth]{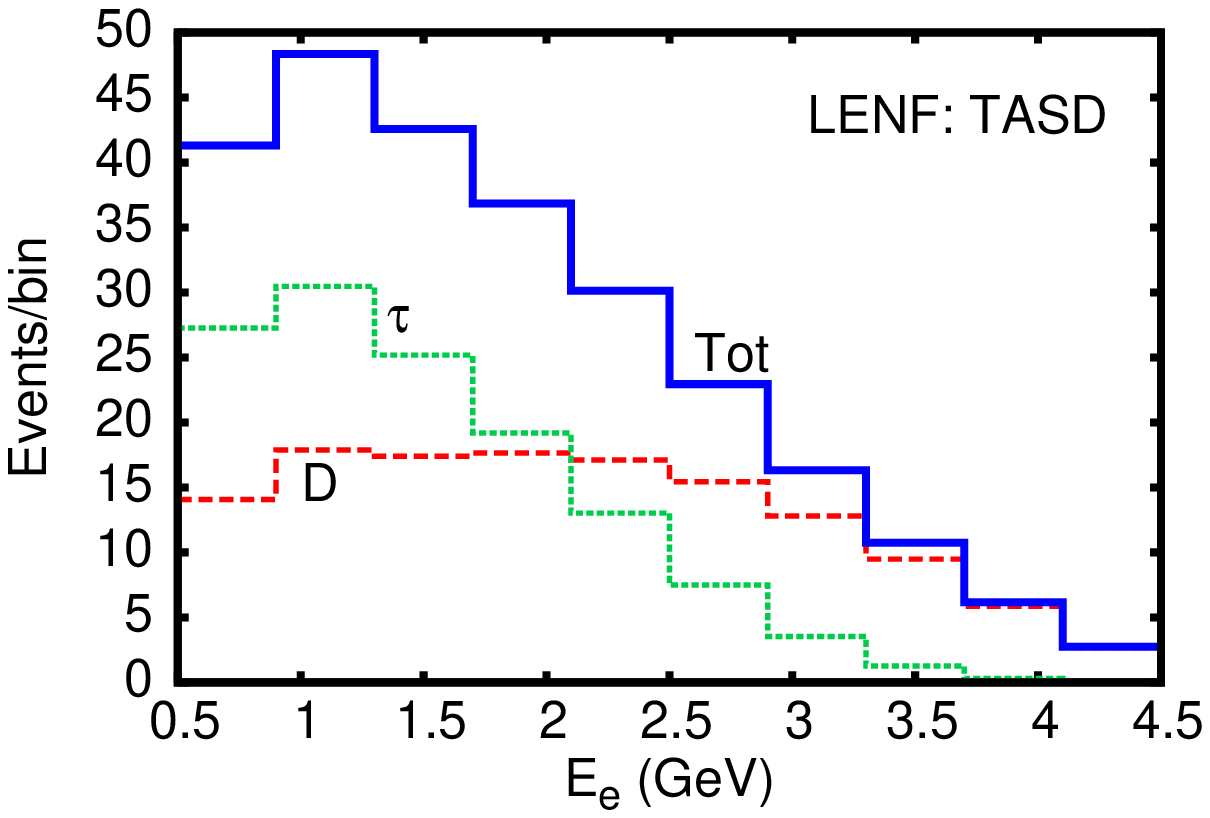}
\includegraphics[width=0.45\textwidth]{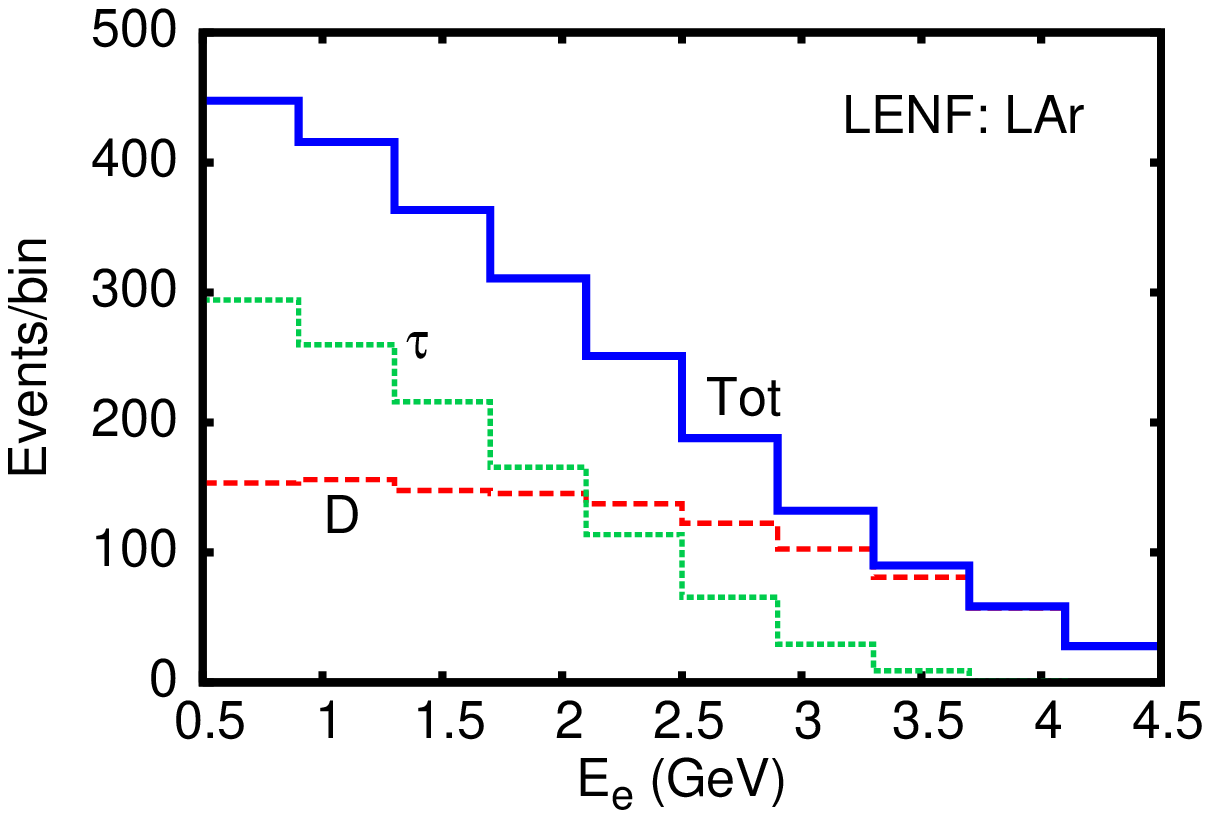}
\caption{Event rates as a function of the observed electron energy at a
TASD detector (L) and LAr detector (R), in the platinum channel alone;
for more details, refer the text. It is seen that the tau-induced events
dominates over the direct WS electron events at low values of the observed
electron energy.}
\label{fig:lenfevents}
\end{center}
\end{figure}

It is obvious from Fig.~\ref{fig:lenfevents} that the tau contribution
(labelled $\tau$) is significant compared to the direct events
(labelled D) even for a low energy neutrino factory. (The difference
in shape of the events at TASD and LAr detectors at low energy is due
to the jump in the electron efficiency above $E_e = 1$ GeV for the
TASD; however, the tau-induced events are a sizeable fraction in
both). This is a surprise, and can be understood from the fact that
the $\nu_\tau$'s arise dominantly from $\nu_\mu \to \nu_\tau$
oscillations and that the $\nu_\mu$ spectrum peaks near the parent
muon beam energy. Hence a substantial fraction of the $\nu_\tau$'s
have sufficient energy, $E > E_\nu^{thr}\sim 3.4$ GeV, to produce taus
in the detector. These taus decay preferentially into low energy
leptons \cite{IN}. Thus one needs to take into account the
contribution coming from the tau neutrinos to the total WS electron
events (labelled ``Tot" in the figure) in order to obtain the correct
constraint on the oscillation parameters such as $\theta_{13}$ and
$\delta_{CP}$.

This situation is in contrast to the tau contribution to the muon sector,
where these dominant $\nu_\tau$'s contribute to the muon disappearance
(RS) events. The tau contribution to the golden channel (WS events)
arise from the highly suppressed $\nu_e \to \nu_\tau$ oscillations; tau
production is further suppressed at a LENF since the $\nu_e$ spectrum
peaks at about only two-third of the parent muon beam energy, so that
relatively fewer $\nu_\tau$'s have energies larger than the threshold
energy for CC tau interaction.

\begin{figure}[btp]
\begin{center}
\includegraphics[width=0.45\textwidth]{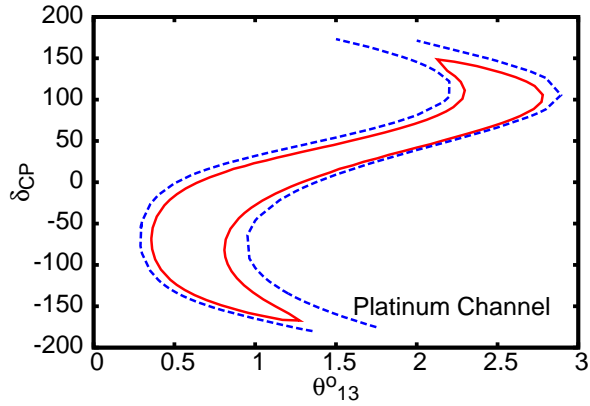}
\caption{99\% CL contours with (outer) and without (inner curve) inclusion
of tau-induced electron events as in Fig.~\ref{fig:henfsens} with platinum
channel alone, for a LENF with TASD detector at a baseline of $L=1300$
km and 10 years running time.}
\label{fig:lenftasdsens}
\end{center}
\end{figure}

The corresponding 99\% CL allowed contours in $\theta_{13}$--$\delta_{CP}$
parameter space are shown for the same input parameters, for the TASD
detector, in Fig.~\ref{fig:lenftasdsens}.  While $\theta_{13}$ can
still be discriminated from zero, in this case again, there is little
sensitivity to $\delta_{CP}$ which is worsened by the tau contribution.

Again inclusion of backgrounds will further worsen the sensitivity
to the various channels; in fact, the inclusion of tau-induced events
can be considered as an additional background to the main signal; this
can seriously limit the efficacy of the platinum channel, as has been
discussed in Ref.~\cite{whichlenf}.

\begin{figure}[htp]
\begin{center}
\includegraphics[width=0.45\textwidth]{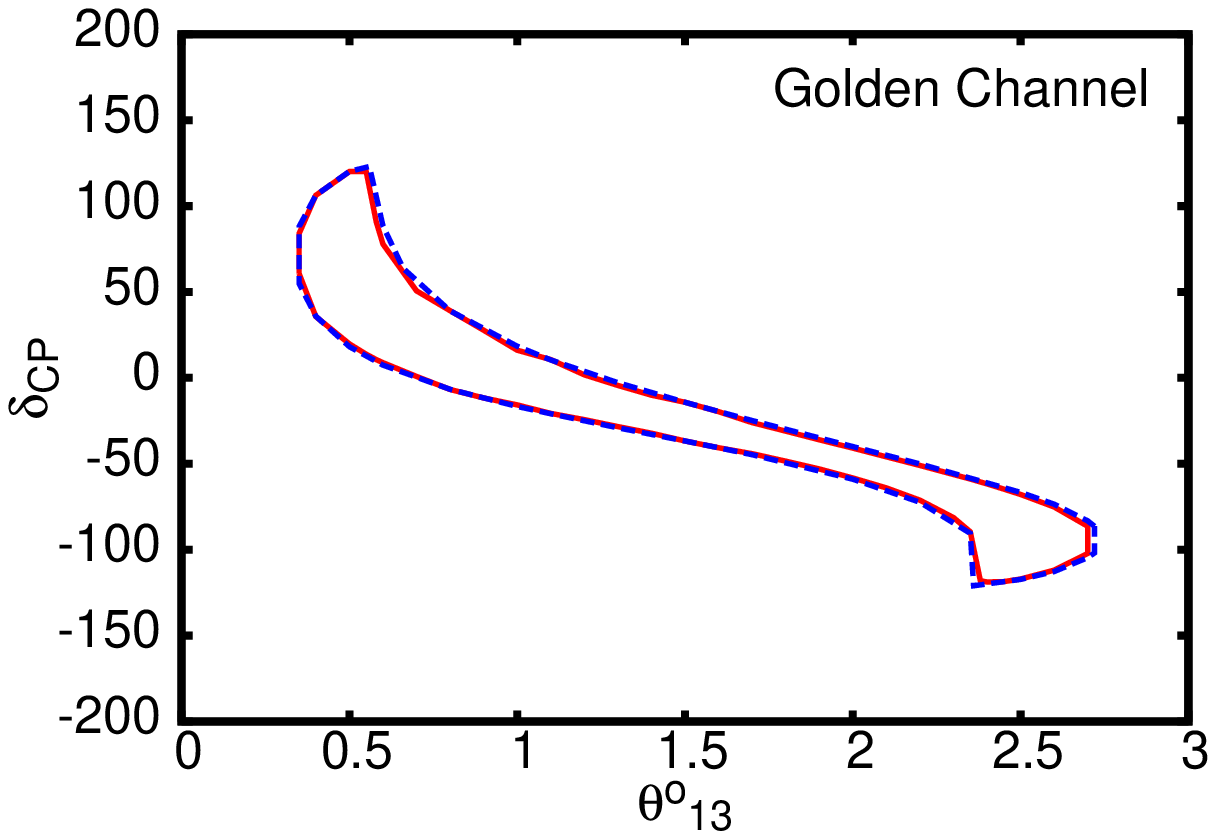}
\caption{99\% CL contours with and without inclusion of tau-induced
muon events in the golden channel alone (WS muons), for the same
oscillation parameter inputs, at a LENF with TASD detector.}
\label{fig:lenftasdgolden}
\end{center}
\end{figure}

Recall that the tau-induced events add dominantly to RS muon events, in
contrast to their substantial contribution to WS electron events. This
is seen by the relatively small sensitivity to tau events of the {\em
golden} channel in a TASD detector in Fig.~\ref{fig:lenftasdgolden}.
Here we have used 73\% (94\%) muon efficiencies below (above) 1 GeV muon
energy with 10\% muon energy resolution \cite{Li}.

\begin{figure}[htp]
\begin{center}
\includegraphics[width=0.45\textwidth]{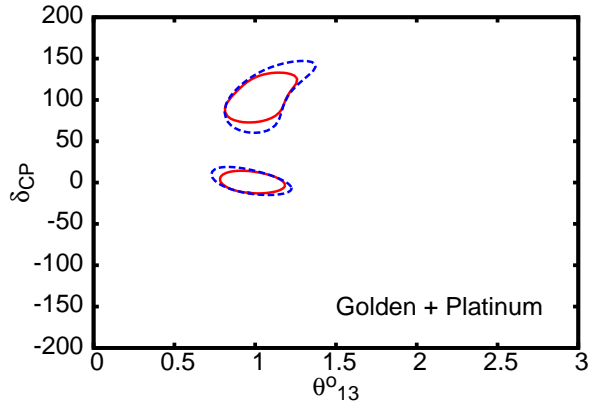}
\caption{99\% CL contours with (outer) and without (inner curve)
inclusion of tau-induced muon events from the combined electron and
muon wrong sign events, or combined golden and platinum channels, for the
oscillation parameter inputs, $\theta_{13} = 1^\circ$ and $\delta_{CP}
= 0^\circ, 90^\circ$, at a LENF with TASD detector.}
\label{fig:lenftasdptau}
\end{center}
\end{figure}

\begin{figure}[htp]
\begin{center}
\includegraphics[width=0.45\textwidth]{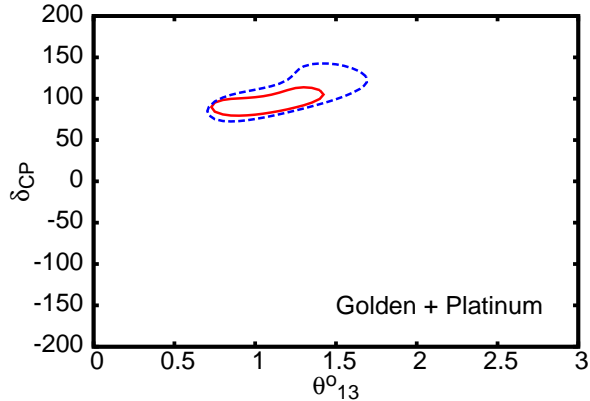}
\caption{As in Fig.~\ref{fig:lenftasdptau}, for a LENF with LAr
detector, with inputs $\theta_{13} = 1^\circ$ and $\delta_{CP}
= 90^\circ$.}
\label{fig:lenflarptau}
\end{center}
\end{figure}

Due to the very different sensitivities of these two channels to the
oscillation parameters, the combined `golden $+$ platinum' channels are
very sensitive to $\theta_{13}$ and $\delta_{CP}$, as can be seen from
Fig.~\ref{fig:lenftasdptau}. The inclusion of the tau-induced
events clearly worsens the sensitivity to both parameters, especially at
the CP-odd point, $\delta_{CP}^{\rm input} = 90^\circ$. This is also
true for the LAr detector, as shown in Fig.~\ref{fig:lenflarptau} for
this set of input parameter values.

\begin{figure}[htp]
\begin{center}
\includegraphics[width=0.45\textwidth]{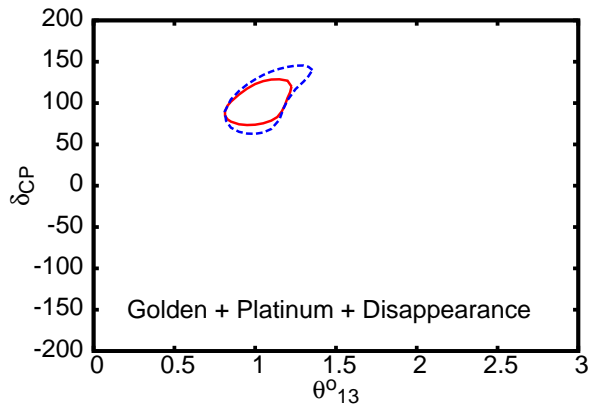}
\caption{As in Fig.~\ref{fig:lenftasdptau}, for a LENF with TASD
detector, on combining the golden and platinum appearance channels (muon
and electron WS events) as well as the muon disappearance (muon RS)
channel.}
\label{fig:lenftasdptaurs}
\end{center}
\end{figure}

Finally, Fig.~\ref{fig:lenftasdptaurs} shows the sensitivity using the
combined data from the golden and platinum channel as well as from
muon disappearance (RS muons). The more the number of channels, the
tighter the constraints expected; while this is true for the `direct
events' alone, the tau-induced events continue to worsen the
sensitivity to these parameters in all cases. This situation is
somewhat ameliorated but still not completely compensated by the
combined analysis of golden and platinum channels, or by the
additional inclusion of muon disappearance events.

A particularly vexing question is that of the corresponding $\nu_\tau$
CC cross-section in the detector. This will remain unconstrained even
with the presence of a near detector which will measure only CC $\nu_e$
and $\nu_\mu$ cross-sections. As long as the normalization uncertainties
associated with the tau channel remain large, the ultimate reach of
neutrino factories for these oscillation parameters, particularly for
$\theta_{13}$ and $\delta_{CP}$ with the platinum channel, will remain
limited. There are proposals to measure the $\nu_\tau$ CC interaction
cross-sections \cite{JC}; this will be a crucial input to reduce
the systematic uncertainties of the $\tau$-induced events.

\section{Conclusion}
\label{CON}

Precision measurements of the oscillation parameters are the main goal
of future advanced neutrino oscillation experiments. Neutrino
factories have particularly good sensitivity to parameters such as the
1--3 mixing angle $\theta_{13}$, the Dirac CP phase $\delta_{CP}$, and
the neutrino mass hierarchy, especially when $\theta_{13}$ is small
and inaccessible at current or near-future reactor and short-baseline
accelerator experiments. While the golden channel (observation of
wrong sign muons) is best suited for these measurements, degeneracies
and correlations worsen the sensitivities obtainable. These can be
lifted by a judicious choice of baselines, improved statistics,
as well as inclusion of wrong sign electron events---the so-called
platinum channel---and has extensively been discussed in the literature
\cite{HA,whichlenf}.

However, muon neutrinos in such factory fluxes can oscillate to tau
neutrinos, driven by the relatively large mixing angle $\theta_{23}$,
which is nearly maximal. These $\nu_\tau$ produce taus in CC
interactions in a detector, which can promptly decay into muons or
electrons (each 17\% of the time). The golden channel is relatively
insensitive to the tau contribution since taus contribute dominantly
to muon {\em right sign} events \cite{IN} and hence do not
substantially spoil the results in this sector \cite{HA, Donini}. This
paper studies for the first time the impact of the contribution that
comes from decay of taus to the total electron wrong sign events
(platinum channel) in a detector. Studies of both high energy (25 GeV
muon beam) as well as low energy (4.5 GeV) neutrino factories reveal
that the tau contribution is substantial and alters (in fact worsens)
the sensitivity to precision measurements of $\theta_{13}$ and
$\delta_{CP}$.
Uncertainties from the tau background in all the channels must be brought under control to obtain precision measurements at neutrino factories.


\end{document}